\documentclass[]{article}
\usepackage{lmodern}
\usepackage{amssymb,amsmath}
\usepackage{ifxetex,ifluatex}
\usepackage{fixltx2e} 
\ifnum 0\ifxetex 1\fi\ifluatex 1\fi=0 
  \usepackage[T1]{fontenc}
  \usepackage[utf8]{inputenc}
\else 
  \ifxetex
    \usepackage{mathspec}
  \else
    \usepackage{fontspec}
  \fi
  \defaultfontfeatures{Ligatures=TeX,Scale=MatchLowercase}
\fi
\IfFileExists{upquote.sty}{\usepackage{upquote}}{}
\IfFileExists{microtype.sty}{%
\usepackage{microtype}
\UseMicrotypeSet[protrusion]{basicmath} 
}{}
\usepackage{hyperref}
\hypersetup{unicode=true,
            pdfborder={0 0 0},
            breaklinks=true}
\usepackage{longtable,booktabs}

\ifx\paragraph\undefined\else
\let\oldparagraph\paragraph
\renewcommand{\paragraph}[1]{\oldparagraph{#1}\mbox{}}
\fi
\ifx\subparagraph\undefined\else
\let\oldsubparagraph\subparagraph
\renewcommand{\subparagraph}[1]{\oldsubparagraph{#1}\mbox{}}
\fi

\usepackage{makecell}
\usepackage{hanging}

\date{August 17, 2005; revised December 20, 2005 \& December 21, 2006; reformatted June 29, 2016}

\title{\textbf{Practical Testing of a C99 Compiler Using Output
Comparison}}
\author{Flash Sheridan \\ Quality Lead for Compilers \\Tools Quality Assurance \\ \url{http://pobox.com/~flash} \\ Access Systems Americas, Inc. (formerly PalmSource, Inc.) \\ 1188 East Arques Avenue \\ Sunnyvale, CA 94305-4602}

\begin{document}
\maketitle

This is a preprint of an article accepted for publication in

\emph{Software: Practice and Experience}

Copyright © 2002-2006, Access Systems Americas, Inc.

\begin{abstract}
A simple technique is presented for testing a C99 compiler, by
comparison of its output with output from preexisting tools. The
advantage to this approach is that new test cases can be added in bulk
from existing sources, reducing the need for in-depth investigation of
correctness issues, and for creating new test code by hand. This
technique was used in testing the PalmSource Palm OS® Cobalt ARM C/C++
cross-compiler for Palm-Powered® personal digital assistants, primarily
for standards-compliance and correct execution of generated code. The
technique described here found several hundred bugs, mostly in our
in-house code, but also in longstanding high-quality front- and back-end
code from Edison Design Group and Apogee Software. It also found
eighteen bugs in the GNU C compiler, as well as a bug specific to the
Apple version of GCC, a bug specific to the Suse version of GCC, and a
dozen bugs in versions of GCC for the ARM processor, several of them
critical.
\end{abstract}

\\

\section{\emph{Introduction}}

The most time-consuming part of creating tests for a compiler is writing
new code for new test cases, and deciding whether these cases should
pass or fail, which requires detailed language-lawyer skills. This paper
presents a simple technique for avoiding such effort in a large number
of cases, provided that an existing tool or tools are available for
comparison of output.

The basic insight is that different compilers are likely to have
different bugs, and so bugs in one compiler are likely to result in
output which differs from another compiler's for the same input. The
preexisting tools do not need to accept exactly the same language as the
tools under test, nor run on the same platform. Nor does their quality
need to be perfect, though bugs in the preexisting tool are likely to
look initially like bugs in the tool under test.\footnote{Indeed, the
  compiler under test incidentally found four bugs in GCC {[}1, GCC{]}.
  Testing of our assembler using these techniques found five bugs in the
  ARM assembler and/or its documentation.} All that is required is a
substantial overlap between the two tools, and a reasonable level of
quality in the reference tools. Two forms of output need to be
distinguished here: for simple tools (such as an assembler) comparison
of the output of the tool itself with the output of the reference tool
may suffice. For a more complex tool, such as a compiler, it is instead
generally necessary to compare the output which results from executing
the output of the tool.

The techniques presented here were developed for testing the PalmSource
Cobalt ARM C/C++ embedded cross-compiler, part of the Palm OS Developer
Suite for Windows {[}2, Palm OS Developer Suite{]}, used to build
applications for personal digital assistants running Palm OS® Cobalt
{[}3, Palm OS® Cobalt{]}. (Very few devices running Cobalt were
released; PalmSource switched to the Linux platform soon after testing
of this compiler was completed.) The main goal of testing was ensuring
the correct execution of the machine code generated by the compiler; the
secondary goal was conformance to language standards and our own public
documentation. The two reference tools used for output comparison with
the PalmSource compiler chain were the Gnu C Compiler {[}4.5=1, GCC{]}
in C99 {[}4, C99{]} mode (primarily version 3.3, with occasional
reference to 3.4 later in the project), and the ARM ADS {[}5, ADS{]} 1.2
assembler. PalmSource's compiler used licensed code from Apogee Software
{[}6, Apogee{]} for the back-end, which in turn used a front-end from
the Edison Design Group {[}7, Edison Design Group{]}. A licensee of
Apogee's code needs to provide additional target-specific compiler code,
plus an assembler. The focus of PalmSource's testing was on our own new
code.

The tests found three hundred and fifty-three bugs in our compiler,
mostly in the code we had written in-house, but also found bugs in
Apogee and Edison code. Both of the latter are long established and well
respected, which suggests that compiler vendors could profitably use the
simple techniques presented here. Of course, this technique presents
some risk of missing bugs which occur in identical forms in both the
reference compiler and the compiler under test. No testing strategy for
a commonly-used language is likely to eliminate all bugs, and such bugs,
at least, are unlikely to be very prominent, since corresponding bugs
presumably shipped in a known product of reasonable quality.

During the course of this project, at the insistence of PalmSource's
third-party developers, official support for C++98 {[}8, C++98
ISO/IEC{]} was added to the product. Additional testing for C++ was
handled by a different tester, so this article largely focuses on C99
testing, though of course much of the testing for C99 also applied to
C++ code. Since the completion of this project and PalmSource's switch
to Linux, we have been using the tools discussed in this paper in
reverse, testing various versions of GCC by comparison of its output
with our own compiler.

The techniques presented here are a supplement to existing test suites,
not a replacement. GCC has its own test suite, which is discussed below;
obviously making use of it, if practical, would provide valuable
testing. But the simple techniques presented here found bugs missed by
GCC's testing, in the Gnu {[}9, GCC Bug Database, bugs 15549, 15642,
21613, 21659, 22508, 22603, 22604, 23089, 23118, 23125, 23223, 23225,
23307, 24075, 24322, 26774, 26818, 26865, 27829, 27851, 27881{]}, Apple
{[}10, Apple Bug Database bug 666675{]}, Suse Linux {[}11=9, GCC Bug
23472{]}, and ARM {[}12=9, GCC bugs 24075, 24322, 27829; 13,
CodeSourcery arm-gnu mailing list messages 346 and 381; CodeSourcery bug
database bugs 401, 440, 444, 456, 458, 502, 726{]} versions. GCC's test
suite is extensive and complicated; I believe that these results show
that simpler testing for it would be worthwhile as well.

Commercial conformance suites are also available, and any organization
undertaking to produce a compiler would be well advised to purchase one,
but a nondisclosure agreement precludes discussing the limitations of
the test suite which PalmSource used. Any new compiler will require
validation well beyond mere conformance testing; the obvious temptation
is to try to write all such cases oneself, deciding \emph{a priori} what
the results of each case should be. This cannot be avoided entirely, and
writing some high-level language feature tests in-house is likely to be
important, but this should be done only where writing new code is truly
necessary, generally for features specific to the new compiler.

\section{\emph{General Considerations on Compiler Testing}}

Testing a compiler is, above all, a software testing endeavor; two of
the standard works on testing are {[}14, Kaner et al{]} and (though much
of it is from the mainframe era) {[}15, Beizer{]}. Given the complexity
of the system of which a new compiler is likely to be a part, the
conventional wisdom applies even more than usual: First, ensure that
there is an adequate specification for the entire toolchain of which the
compiler is a part, \emph{before} testing starts. Second, ensure that
the high-level specification matches reality. Much more than user-level
software, the quality of a developer tool is largely determined by the
correspondence between the public specification and the actual behavior.
There will be many compiler bugs which can be resolved just as well by
changing the documentation as by changing the behavior. Such issues are
surprisingly easy to miss, since they can be taken for granted by all of
those familiar with the project from its early days.

Therefore, test breadth-first, not depth-first. It can be very tempting
to test an interesting detail in depth, but in our experience, this
sometimes led to missing much higher-level and more obvious bugs, which
would have had a more severe impact on end users. More generally, it is
woefully common in software testing to spend a great deal of time on
difficult tests, finding obscure bugs, while missing more obvious and
important ones. This phenomenon suggests an important law of quality
assurance: \emph{Nothing is ever too obvious to need systematic
testing.}

Three of the major compiler testing variables are optimization, debug
information in the output code, and debug checking within the compiler
itself. A good compiler will have a debugging version which
double-checks itself as it compiles, terminating with an assertion
failure if something goes wrong (GCC activates this when it is built
with the -\/-enable-checking=all option). Testing such a version can be
extremely useful; especially for a cross-compiler, it is much easier to
elicit bugs via a compile-time abnormal termination than through
run-time incorrect execution. Note that this is distinct from debug
\emph{output} produced by the compiler; most compiler vendors ship a
non-debug version of their compiler (though a preliminary limited
release of a debug version is a good idea, and testing with it
internally is crucial), but even the release version of the compiler
must be tested for the production of debug output.

Rather unusually for quality assurance, crashes for a compiler are
relatively benign, at least compared to execution-correctness bugs. A
compiler crash or assertion failure affects only the programmer, and
will be immediately obvious to him or her. The programmer may be able to
work around the problem, and if the problem is reproducible, it will
probably be straightforward to fix the compiler. In our experience, such
crashes usually occurred, at least after the initial testing, when
exercising relatively obscure compiler functionality, and hence were
unlikely to affect a large number of programmers.

A code-correctness bug, on the other hand, may not be immediately
apparent to the programmer. Ideally it will be noticed by testers, but
it is possible (especially if the bug occurs only at high optimization)
that it will only be noticed by end users after a product built with the
compiler has shipped. This could require a recall of products built with
the buggy compiler and sold to a large number of end-users.

As well as testing with both debug and release versions of a compiler,
testing is required for the production and omission of debug output, and
various levels of optimization in generated code. Any given compiler
test could theoretically be run at all combinations of optimization
level and debug output.\footnote{Our compiler had five levels of
  optimization and, in addition to the usual options to produce debug
  output, or to produce no debug output, a third option which produced
  debug output, while still allowing the inlining of functions.} In our
experience, when time was short, the most useful option was to test with
full optimization, and either of the debug-output options. A better
compromise was to test with all three debugging options, and with
minimal and maximal optimization, making six combinations. Bugs specific
to other cases were quite rare, though there were a few files which
produced warnings only at intermediate optimizations. This was not a
bug, though it initially looked like one: The compiler simply didn't
notice the issue without optimization, and at the highest optimization,
greater analysis revealed that what looked like a potential
uninitialized reference could in fact never happen.

\section{\emph{I: Compiler Acceptance/Rejection/Crash Testing}}

The first, and simplest, technique is testing simply for correct
compiler acceptance or rejection of individual source files. (A related
issue was the issuance of warnings.) This had the side benefit of crash
testing the compiler, especially at high optimizations, which turned out
to be more valuable than the acceptance/rejection testing. The key
observation was that, while GCC's C99 support was not perfect, any C
file which it accepted in C99 mode, and which our compiler also accepted
without optimization, was likely to be a valid test case. Any file which
both compilers rejected was likely to be a valid rejection test case.

A fortunate circumstance was that GCC also ships an extensive test
suite, the GCC C-Torture Test Suite {[}16, GCC C-Torture{]}, with a
large number of test source files. Many of the C files were not valid
C99; the most common problem was failure to prototype a function or to
include the header for it, most frequently printf. However, there were
over a thousand files (slightly more than half of the total) which were
valid C99, and accepted both by GCC in C99 mode and by our compiler.

It was straightforward to write a Perl script which recursed the entire
C-Torture source hierarchy, recording which files were accepted by each
compiler, and which were rejected. An \emph{ad-hoc} script indicated
which files were supposed to be rejected (or to generate warnings), by
moving them into appropriately-named subdirectories. The files on which
GCC/C99 and our compiler (without optimization) agreed, were checked
into our source repository. A second Perl script then recursed over the
checked-in files, calling our compiler at every optimization and debug
level, on each supported platform, for each build.\footnote{For
  non-milestone builds, the recursion was limited to the highest and
  lowest optimization, since we had very few bugs which happened only at
  intermediate levels of optimization, but many bugs which occurred only
  at the highest optimization levels.} This found a few unexpected
acceptances (or rejections) at higher optimizations, or on different
platforms; normally acceptance should not be affected by optimization,
and cases in which this happens deserve investigation. One such case
turned out to be a genuine compiler bug; another case involved a linker
error caused by an undefined reference, which was optimized out of
existence at higher optimizations. A number of third-party source files
showed such discrepancies without revealing bugs, most frequently
because of invalid assembly code, and less frequently because of code
explicitly conditionalized on optimization. (There is an addition
possibility with some versions of GCC, which will reject some invalid
code, e.g., taking the address of a cast expression, only at higher
optimizations.) The greater benefit of this technique was compiler
crashes or debug errors at higher optimizations. The GCC C-Torture files
were very good at eliciting such bugs, but once the technique was set
up, it was straightforward to add additional files from other sources,
such as sample code available on the web, or the source files for Palm
OS®. In one instance, we made a test case out of a complaint from a
language purist in an internal chat room, who was annoyed that a rival
compiler was willing to accept obviously non-standard code.

This machinery could also be applied for crash testing with arbitrary
C/C++ code. This simply required pointing the recursive script at a
directory of source code, and disregarding the accept/reject results,
focusing solely on reported crashes or debug assertions. Open source
projects were a good source of C/C++ source files for crash testing. We
recursed over the source for Cygwin {[}17, Cygwin{]}, Perl {[}18,
Perl{]}, Blitz++ {[}19, Blitz++{]}, and GCC {[}20=1, GCC{]}, as well as
our internal source tree, but the most fruitful source of compiler
crashes and errors were GCC C-Torture {[}21=16, GCC C-Torture{]} and the
Boost C++ libraries {[}22, Boost{]}. Once a crash is found and
reproduced, a tool such as Delta {[}23, Delta; 24, Zeller{]} should be
used to reduce the code necessary to elicit the crash; this is
especially important if the bug is to be reported publicly, but the code
causing the crash is confidential. A Windows runtime check also found an
uninitialized variable when the compiler was trying to compile itself;
this is not a realistic test scenario for an embedded compiler, but
turned out to be an interesting stress test.

Execution-correctness testing (covered in the next section) for such
projects would have required getting them to execute under Palm OS,
which would have been a significant engineering effort. Crash testing
during the compilation phase is a cruder and easier technique than
testing run-time execution, but given effective debugging checks in the
compiler, can be surprisingly effective. One of the more interesting
crashes found by this technique occurred only at optimization level 2 or
higher, compiling the GCC C-Torture file 20010404-1.c, whose source code
is below. Coincidentally, this file was added to the GCC test suite
because it caused a crash in GCC at the same optimization level. The bug
in our compiler was that constant-folding optimization asked the host
processor to do arithmetic without checking the arguments for validity.
This converted a run-time error (which is acceptable for invalid
arguments) into a compile-time crash, which is not. More generally,
dividing the most negative value of a signed integer type by -1 is a
fruitful source of bugs, since it will overflow if integers are
represented in twos-complement form.\footnote{{[}25, Harbison \&
  Steele{]}, page 228.}

\bigskip
\begin{tt}
\#include \textless{}limits.h\textgreater{}

extern void bar (int);

void foo ()

\{

  int a = INT\_MIN;

  int b = -1;

  bar (a / b);

  bar (a \% b);
 
\samepage
\}
\end{tt}

\emph{Figure I: GCC C-Torture file 20010404-1.c}
\bigskip

Note that the compile-time performance of arithmetic, which here
elicited a bug, can also mask problems. When testing run-time
operations, it is often necessary to mark variables as volatile, to
prevent optimization from removing an operation from the generated code.

\section{\emph{II: Emulated-Execution Output Correctness Verification}}

Our principal technique for validating correct execution of
generated-code used the output from a cross-platform desktop ARM
emulator, based on an open source ARM emulator {[}26, ARMulator{]} but
with significant in-house modifications. (Note that this is the second
form of output correctness validation mentioned above: verifying the
correctness of the output from the execution of the output of the
compiler.) Some such emulator is useful for testing a cross-compiler
targeted at an embedded device, though ours had significant limitations:
Its runtime environment was fundamentally different from a real
device's, which allowed the possibility of device-only compiler bugs.
(It also lacked a graphical user interface, which precluded its use by
application developers, but was less of a limitation for compiler
testing.) If at all possible, anyone involved with a cross-compiler
should insist on the importance of an emulator whose runtime environment
(and, preferably, graphical user interface) matches the actual product
closely. Even a limited emulator, however, was useful for broad coverage
of basic code-correctness issues, using any available
standards-compliant source code.

The method for taking a standards-compliant source code file, and
converting it to a code-correctness test case, was simple but limited.
Once a suitable source file was identified, a Perl script called GCC on
the source file, and then ran the resultant desktop native executable.
The script captured the output, and appended it to the source file,
wrapped in comment markers. Once the file was supplemented with its
expected output, and added to the test list, a different script would
call our compiler on it (at some or all optimizations). This script
would then run the executable under the desktop emulator, comparing the
actual output to the expected, and reporting the first difference if the
output was not as expected.

One advantage of the simplicity of this approach to compiler testing is
that a wide variety of source files are suitable for execution
comparison testing. Provided only that a file's main purpose was to
write to the standard output, and it was careful about including the
standard headers it required, and did not rely on anything outside the
C99-mandated libraries, then it was a plausible candidate for addition
to the test suite. Some source files' output depended on
platform-specific details, such as the size of data types. This could be
a serious difficulty for testing a multi-target compiler, but since our
compiler had only a single target, it was usually straightforward to
work around. Somewhat surprisingly, Macintosh OSX {[}27, Macintosh
OSX{]} turned out to be the most convenient desktop environment for
generating output for comparison, as its run-time environment had much
in common with the ARM Application Binary Interface {[}28, ARM ABI{]}
which we were targeting.

A source file can even be useful if it only prints on failure, as long
as its error reporting is reasonably specific. Specificity of failure
reporting is even more important than usual for compiler correctness
testing: At all but the earliest stages of development, a large majority
of apparent test failures are likely to be bugs in the test rather than
in the compiler.

A more advanced technique was to generate standard-compliant C99 or
C++98 code automatically, using Perl. A script rotated through a list of
constants of the principal arithmetic types, producing a source file for
each of a list of operators, printing out the result of the operator for
each pair of constants. The initial intent was to choose one number from
each test case equivalence class,\footnote{{[}29=14, Kaner et al.{]},
  page 125--132.} but this quickly generated impractically large files
for binary operators, so a somewhat arbitrary subset of numbers which
seemed likely to elicit bugs was used instead. This found only a single
bug in our compiler, which had already been found by test code generated
from the documentation; the low bug count was probably because this code
was primarily testing high-level code paths implemented by Edison and
Apogee. It also found two bugs in Gnu GCC {[}30=9, GCC bugs \#15549 and
\#24075{]}, one bug in Suse Linux's version of GCC {[}31=9, GCC Bug
23472{]}, and five bugs in CodeSourcery's version of GCC for the ARM
processor {[}32=13, CodeSourcery; 33, CodeSourcery 2005Q3-2 Release
announcement, issues 1 \& 3{]}. This technique also elicited some
unfortunate behavior on reasonable but out-of-spec code {[}34=9, GCC bug
\#16031{]}, converting a floating-point number to a long long integer.
The last issue showed the need to exclude out-of-spec code from being
generated. Doing this in full generality would have required evaluating
C expressions in Perl, but simple tweaking of the lists of constants,
plus some special-case exclusion code, sufficed for our purposes.

This comparison technique also made it simple to validate
manually-created expected output. A particularly fruitful approach was
to take tables of constants from the public documentation, and massage
them in a text processor into both print statements and expected output.

This technique provided reassurance that optimization, debug
information, host platform, and endianness issues did not affect code
correctness, at least under emulation. It found gratifyingly few
code-correctness bugs, mostly in the desktop emulation environment.
Given the criticality of any code correctness issue in a widely
distributed compiler, testing like this is important even if it finds
few of the bugs it targets. It found considerably more compiler crash or
debug-assertion failures; as previously noted, these (though serious)
were considerably less threatening than code-correctness bugs. One of
the most interesting code-correctness bugs this technique found was the
above-mentioned bug in GCC {[}35=9, GCC bug \#15549{]}, though our
script initially reported it as a bug in our compiler. The symptom was
that the (admittedly rather odd) arithmetic comparison ``true
\textless{} \textsf{`}a\textsf{`}'' returned false rather than true. This comparison was
part of a large number of automatically generated operations, mentioned
above. The underlying cause was that GCC used a simplified
implementation of the ``less than'' operator when one of the arguments
was a Boolean value, which was invalid unless both operands were
Boolean.

The following were useful sources of standard-compliant code for
emulated execution: C-Torture {[}36=16, GCC C-Torture{]} has been
mentioned above; all of its files were useful for crash testing, and a
subset were C99-clean and useful for execution testing using our
framework.\footnote{Actual usage of the C-Torture suite framework itself
  was impractical for our compiler testing, not because of problems with
  the GCC suite itself, but because of problems with the underlying
  framework, DejaGnu. {[}52, DejaGnu bug mailing list message 2006-02/3
  et al.{]}} Lindig's Quest test system {[}37, Lindig{]} was a good
source of clean code for argument-passing tests, though it fortunately
failed to find any bugs in our compiler, presumably due to the quality
of Apogee's and EDG's code; it has found a number of bugs in other
compilers, and is recommended for new compiler testing. Tydeman's
mathematical tests {[}38, Tydeman{]} and the Paranoia floating point
test suite {[}39, Kahan et al.{]} were useful for testing compiler
correctness, though their focus is more on the mathematics library than
on the compiler. The source code for the O'Reilly \emph{C++ in a
Nutshell} guide {[}40, Lischner{]} has a number of useful C++ tests. The
STLPort {[}41, STLPort{]} self-test was a very good test of the compiler
in general, without needing any of the machinery presented here; it
tests run-time correctness as well as compile-time acceptance of valid
code. STLPort is an implementation of C++ standard libraries, but it
exercised portions of the compiler which were relevant to both C and
C++.

lcc {[}42, lcc; 43 Fraser \& Hanson{]} ships with a test suite
containing approximately seven thousand lines of new code, not counting
the Paranoia suite, which the suite also includes. The most notable test
is a file designed to have one test case for each section of the
language specification in {[}44, Kernighan \& Ritchie{]}. Unfortunately
the test code relies on behavior deprecated in ANSI C, and will be
rejected by a strict C99 or C++98 compiler; it can be useful, however,
for testing a compiler which can accept old-fashioned C. SDCC {[}45,
SDCC{]} includes approximately eight thousand lines of test code; the
bulk of the source files are reasonably portable; but some crucial
included code is highly non-portable, and would need to be adjusted to
run on another compiler, and the error reporting is extremely terse.

\subsection{Device Execution Correctness Testing}

Code-correctness on an emulator is, of course, much less important than
correctness on the device; but building for, and running on, the device
was much more time consuming, and not subject to automation. Typically
we ran the emulated execution script for each build on each supported
platform, at the highest and lowest optimizations, with tests at all
optimizations for major milestones. The latter took more time, though
since it could run unattended, this was not a major concern.

Converting the test code to run on the device required another script
and substantial manual intervention. The script iterated over the test
list, skipping special cases and expected failures. For each test case,
a series of commands was generated to copy the source file from our
source control system to the directory for the device application. A
second iteration of the script then searched the source file,
conditionally changing the definition of ``main()'' to a unique name
based on the source file name. Three other files were also generated: a
header file with declarations of all the former mains, a source file
which called all of the renamed mains in order, and a concatenation of
all the files' expected output.

Considerable manual intervention was required, especially for C files,
to avoid name collisions. C++ files could use namespaces to avoid these
(name collisions for classes had been particularly difficult to
diagnose), but C files often required the manual addition of ``static''
to function definitions to avoid collisions. For C++ streams, additional
care was necessary, since an I/O manipulator, whose effect was
restricted to a single file in the desktop test scripts, would affect
all subsequent C++ files in the combined executable. Once built, the
application was run on the device, and its output compared to the
expected output. As anticipated, the device's behavior largely matched
the emulator's, and hence our compiler also matched GCC's code
generation. This technique found several discrepancies in our compiler's
output, fortunately relatively minor. A similar approach for testing
PalmSource's version of the Freely Distributable Mathematics Library
{[}46, FDLIBM{]} found forty-nine bugs in our modifications, but
fortunately found no bugs in FDLIBM itself. The most interesting of
these bugs was that the value of hypotf was incorrect, due to a
confusion of two registers, in the comparatively rare case that its
arguments differed greatly.

\subsection{Test Harness}

PalmSource makes extensive use of an internal API (Application Program
Interface) test harness, which runs at least one positive and negative
test for each of our publicly-documented functions. Initially this
seemed promising to apply to compiler testing, but it turned out to
require a great deal of effort while finding no compiler bugs. Comparing
the output from the harness's execution for the largest test suite
(including failed test cases), with that from the same source code using
the ADS compiler {[}47=5, ADS{]}, turned up only a single discrepancy in
nine hundred and ninety-five test cases. This turned out to be a bug in
the source code rather than a compiler problem. The test suite was
relying on the value of an uninitialized variable; the code was merely
fortunate with the older compiler. This outcome may be common when using
a large piece of existing code for compiler testing; compiler bugs are
far less common than bugs in source code which haven't been noticed yet.
This insight has been expressed from a different perspective as one of
Mark Jason Dominus's Maxims for Programmers: ``Looking for a compiler
bug is the strategy of \emph{last} resort.'' {[}48, Dominus{]}

\section{\emph{III: Assembler Output Comparison}}

The third technique to be presented found the greatest number of
critical bugs, but is the least general in applicability. It involved
comparing the machine code generated by PalmSource's ARM assembler, with
that generated by ARM's own ADS assembler {[}49=5, ADS{]}. This required
writing a translator to convert the dialect accepted by ADS, to that
accepted by our assembler, which was slightly different. Instructions
which had no simple conversion had to be skipped, which limited the
scope of the testing.

When run on a Windows machine with a valid (and expensive) ADS license,
the PalmSource assembly was translated to ARM assembly, and then
assembled and the relevant machine code saved. This machine code was
then checked into the source control system, so that it was available on
other machines, either Windows machines without an ADS license, or
non-Windows machines, which could not run ADS at all. Thus when the test
was run on a machine without ADS, the output of the PalmSource assembler
on the local machine could be compared with the output produced by ADS
on a different machine. Even simple cases found a number of incorrect
code generation bugs, which surprisingly were not detected by our
commercial conformance test suite. Twenty-one correctness bugs were
found; none of these was of general interest, and were predominantly
transcription errors in the tables of operations and addressing modes. A
non-macro assembler possesses none of the interesting complexity of a
compiler, but even simple transcriptions are subject to error and need
to be tested.

An additional complexity was ignoring the targets of branch
instructions, since ARM's assembler and ours use different relocation
types. This required complicated pattern matching within Perl, run on
the output of Unix or Cygwin diff. (Cygwin {[}50=17, Cygwin{]} contains,
among other things, Windows implementations of some standard Unix tools,
and is extremely helpful for testing on Windows, especially if
cross-platform testing is useful.)

The following sources of assembly were used for comparison. The first
tests were run on each of the examples in the ARM Assembler Guide
{[}51=5, ADS{]}. A second round of testing contained one instance for
each opcode supported by both ARM and PalmSource. A third round of
testing was based on a partial list of opcodes and addressing modes
provided by the software engineer working on the assembler. A fourth
round of test generation was based on the ARM Assembler Guide's language
syntax definition. In retrospect, an off-the-shelf expression generator
should have been used to generate random ARM assembly instructions for
comparison. Instead a Perl script was built incrementally, taking each
ARM instruction definition and generating instances. An advantage,
however, of the incremental in-house approach was that simple
substitutions detected simple bugs, and bug hunting could proceed
incrementally. Some tests were run using assembly generated from the C
source files. Even using only a ``Hello World'' program turned out to be
surprisingly difficult, so testing was largely restricted to
artificially-created assembly code.

\section{\emph{Conclusion}}

The techniques presented here are neither universal nor comprehensive:
Some of them would need to be altered substantially for use on a
different compiler; and other techniques---commercial or Open
Source---would still be needed to supplement them. Variations on these
techniques may be useful for other compilers, especially early on, when
finding simple and easily reproduced bugs is more important than breadth
of test coverage. It is worth getting simple sanity checks up and
running early, long before more complicated tests are ready to run.

The technique of recursive crash testing over the complete GCC C-Torture
suite is straightforward to implement, and was a surprisingly rich
source of optimization bugs; it is probably worth trying for any C/C++
compiler. Acceptance/rejection testing over C-Torture requires
significantly more attention, but should be practical for any compiler
accepting a language close to one of the GCC options. Comparison of the
output from execution of generated code with that from GCC's generated
code requires substantially more machinery, and considerable harmony
between the runtime of the compiler under test and that used by GCC. It
may not be practical for all compilers, but it supplies a great deal of
assurance about code-correctness. Assembler output comparison is useful
only if there is a preexisting assembler to compare with, with a
language easily translatable to the new assembler's, at least in part.
It is an extremely rigorous test, however; so if any component of a tool
chain can be tested in a similar manner, it should be.

Below are four tables summarizing the origins of our test cases and
bugs; the numbers are necessarily approximate and somewhat subjective,
since our bug and source-code databases are designed to help ship a
product, not to record historical information. The test code table is
skewed by the required omission of the confidential commercial
conformance suite we used, as well as by the inclusion of pre-processed
files, which tend to be disproportionately large, but do not necessarily
provide a proportionate degree of test coverage. The absolute numbers of
bugs found are not comparable between tables: The table for our compiler
includes bugs found at all stages of development; the GCC tables include
only bugs found in publicly-released versions.

\bigskip
\emph{Table I: Origins of Test Code (lines of code)}

\begin{longtable}[c]{@{}ll@{}}
\toprule
\textbf{Automatically Generated} & 1,119,302\tabularnewline
\endhead
\textbf{Gnu Code} & 51,660\tabularnewline
\textbf{Boost} & 302,039\tabularnewline
\textbf{Other (Non-Confidential)} & 1,170,235\tabularnewline
\textbf{Total (Non-Confidential)} & 2,643,236\tabularnewline
\bottomrule
\end{longtable}

\bigskip
\pagebreak [4]


\emph{Table II: Origins of PalmSource Compiler Bugs}

\begin{longtable}[c]{@{}l||l|l|l|l|l|l|l||l@{}}
\toprule
& \textbf{\makecell[l]{Assem-\\bler}} & \textbf{\makecell[l]{C Com-\\piler}} & \textbf{\makecell[l]{C++\\ Compiler}} &
\textbf{\makecell[l]{Opti-\\mizer}} & \textbf{Linker} & \textbf{\makecell[l]{Runtime\\ Libraries }} &
\textbf{Other} & \textbf{Total }\tabularnewline
\midrule\hline
\endhead
\textbf{\makecell[l]{Compile-\\Time\\ Error}} & 25 & 58 & 32 & 13 & 6 & 6 & 5 &
145\tabularnewline\hline
\textbf{\makecell[l]{Primary \\ Output \\ Comparison}} & 21 & 3 & 1 & & 0 & 1 & 0 &
26\tabularnewline\hline
\textbf{Run-Time} & 0 & 14 & 12 & 11 & 3 & 9 & 1 & 50\tabularnewline\hline
\textbf{Other} & 1 & 53 & 24 & 6 & 31 & 4 & 13 & 132\tabularnewline\hline\hline
\textbf{Total} & 47 & 128 & 69 & 30 & 40 & 20 & 19 & 353\tabularnewline
\bottomrule
\end{longtable}

\bigskip
\emph{Table III: Origins of x86 and Cross-Platform GCC Bugs}

\begin{longtable}[c]{@{}l||l|l|l|l|l||l@{}}
\toprule
& \textbf{\makecell[l]{C Com-\\piler}} & \textbf{\makecell[l]{C++\\ Compiler}} &
\textbf{\makecell[l]{Opti-\\mizer}}  & \textbf{\makecell[l]{Runtime\\ Libraries }} & \textbf{Other } & \textbf{Total
}\tabularnewline
\midrule\hline
\endhead
\textbf{Compile-Time Error} & 4 & 7 & 3 & & 1 & 15\tabularnewline\hline
\textbf{Run-Time} & 1 & & 1 & & & 2\tabularnewline\hline
\textbf{Other} & & & 1 & & & 1\tabularnewline\hline\hline
\textbf{Total} & 5 & 7 & 5 & & 1 & 18\tabularnewline
\bottomrule
\end{longtable}

\emph{Table IV: Origins of ARM GCC Compiler Bugs}

\begin{longtable}[c]{@{}l||l|l|l|l|l||l@{}}
\toprule
& \textbf{\makecell[l]{C Com-\\piler}} & \textbf{\makecell[l]{C++\\ Compiler}} &
\textbf{\makecell[l]{Opti-\\mizer}}  & \textbf{\makecell[l]{Runtime\\ Libraries }} & \textbf{Other } & \textbf{Total
}\tabularnewline
\midrule\hline
\endhead
\textbf{Compile-Time Error} & 1 & 1 & 3 & & 2 & 7\tabularnewline\hline
\textbf{Run-Time} & 3 & & 1 & 1 & & 5\tabularnewline\hline
\textbf{Other} & & & & & &\tabularnewline\hline\hline
\textbf{Total} & 4 & 1 & 4 & 1 & 2 & 12\tabularnewline
\bottomrule
\end{longtable}

The most striking contrast in the above table is the relative scarcity
of run-time bugs in GCC, especially cross-platform or x86-specific ones.
This is not surprising, given the large number of Open Source projects
which use GCC; every user of one of these projects is implicitly testing
the correctness of the compiler's generated code every time he runs an
executable. The relative scarcity of such bugs does not excuse not
testing for them, however. As noted above, run-time correctness bugs are
the most important sort of compiler bug, and it is striking that the
simple techniques presented here found any at all.

It is unlikely that other compiler developers will be in our exact
situation; but the more general point is to find relevant special cases
for the compiler under test. Don't use only difficult techniques with
general applicability; our most surprising lesson was how much useful
testing could be done with simple, high-level, but
implementation-specific techniques.

Anyone contemplating the release of a compiler should investigate
thoroughly both the GCC test suite and commercial test suites. If an
organization considers the cost of a commercial test suite prohibitive,
it is probably not serious about the resources required to ship a
compiler, and it would be well advised to consider re-using existing
technology instead. A compiler is a vastly complicated piece of code,
with a testing space far larger than normal software, and correctness
standards far higher than end-user applications. There are existing
high-quality compilers, as well as existing compiler frameworks. If at
all possible, an organization should reuse one of them, rather than
doing its own work from scratch: It is hazardous as well as inefficient
to reinvent the wheel.

\section{\emph{Acknowledgments}}

I would like to thank the GNU community for a high-quality compiler to
use for comparison, and a great deal of useful test code. This article
is a small attempt to pay back some of the many benefits PalmSource has
received from the Open Source community. Now that PalmSource has
announced that it will be developing for Linux, we have begun applying
these techniques to testing the GNU C Compiler itself by using the
techniques presented here in reverse, comparing its results to that of
our in-house compiler.

I would also like to thank Kenneth Albanowski for the discussion which
led to one of the key techniques described in this article, and to thank
him, Kevin MacDonell, Jeff Westerinen, the Edison Design Group, Apogee
Software, and CodeSourcery for code whose quality allowed me to focus on
the interesting issues presented here. I am grateful to Rob Stevenson
for his C++ compiler testing, which is largely omitted from this
article, and for comments on an earlier version of it. I would like to
thank John Levine and Mark Mitchell for reviewing the initial version of
this article. I am grateful to the anonymous referees for numerous
insightful and helpful comments. Finally I would like to express my
appreciation to Christian Lindig for helpful criticism, as well as an
impressive argument-passing test suite, which reassuringly failed to
find any bugs in PalmSource's compiler, though it found bugs in a number
of other compilers.

\section{\emph{Bibliography}}

The literature on compiler testing is surprisingly scant. There is
substantial literature on the theoretical design of compilers which
would provably not need testing, but the audience for such work is
largely disjoint from that for the testing of compilers for widely-used
languages which will have a substantial user base. There are also a
number of articles on the automated generation of test code, but given
that there is now a substantial base of real Open Source software, this
is less useful than formerly.

This article, and our testing, was firmly directed towards shipping a
high-quality, but imperfect, compiler which would be of practical use to
the developer community. Producing an inherently bug-free compiler for a
theoretically desirable language was not an option. The goal was to
catch as high a proportion of serious bugs as possible in a useful
compiler for two widely-used languages, C99 and C++98.

The best available bibliography is over a decade old, by Dr. C.J.
Burgess of the University of Bristol; it was a posting to the
comp.compilers Usenet newsgroups, below {[}53, Burgess{]}. Bailey \&
Davidson {[}54, Bailey \& Davidson{]} is an academic article on the
testing of function calls, somewhat similar to Lindig's Quest {[}55=37,
Lindig{]}; it contains the interesting observations that "the
state-of-the-art in compiler testing is inadequate'' (p. 1040), and that
in their experience, the ratio of failed tests to bugs was approximately
one thousand to one (p. 1041). The standard work on compiler theory is
\emph{Compilers: Principles, Techniques and Tools} {[}56, Aho et al{]},
commonly known as the Dragon book. It is a good general introduction,
but had little direct relevance to our testing, except for some extra
caution in including examples of spaghetti code; other standard compiler
texts which were consulted, but did not have significant sections on
testing, are omitted from the bibliography. \emph{A Retargetable C
compiler: Design and Implementation} {[}57=43, Fraser \& Hanson{]}
contains a brief section on the authors' experience with testing their
compiler, with some practical advice on the importance of regression
test cases; difficulties in using lcc's regression tests for other
compilers are discussed above, in the section on emulated-execution
output correctness testing. An updated and alphabetized version of this
bibliography will be made available at
\url{http://pobox.com/~flash/compiler\_testing\_bibliography.html}. 

\bigskip
\begin{hangparas}{.25in}{1}
\begin{raggedright}
1, 4.5, 20: GNU Compiler Collection (GCC) \url{http://gcc.gnu.org}, 1987-.

2: Palm OS® Developer Suite:
\url{http://www.palmos.com/dev/tools/dev_suite.html}, free registration
required, 2004-.

3: Palm OS Cobalt®, \url{http://www.palmos.com/dev/tech/oses/cobalt60.html},
2003-.

4: \emph{C99 ISO/IEC Standard,} INCITS/ISO/IEC 9899-1999, second
edition,
\url{http://webstore.ansi.org/ansidocstore/product.asp?sku=INCITS\%2FISO\%2FIEC+9899-1999},
1999. Commonly referred to as ``ANSI C.''

5, 47, 49, 51: ARM Developer Suite (ADS), version 1.2,
\url{http://www.arm.com/support/ads_faq.html}, 2001-2004.

6: Apogee Software, Inc., \url{http://www.apogee.com/compilers.html}, 1988-.

7: Edison Design Group, Inc., \url{http://www.edg.com/cpp.html}, 1991-.

8: \emph{C++98 ISO/IEC Standard,} ANSI/ISO 14882:1998(E),
\url{http://www.techstreet.com/cgi-bin/detail?product_id=1143945}, 1998.

9, 11, 12, 30, 31, 34, 35: GCC Bugzilla Database:
\url{http://gcc.gnu.org/bugzilla}. See also
\url{http://pobox.com/~flash/FlashsOpenSourceBugReports.html}
for an ongoing list of bugs found with the techniques presented here.

10: Apple Bug Database \url{https://bugreport.apple.com/}, free registration
required.

13, 32: CodeSourcery GNU Toolchain for ARM Processors,
\url{http://www.codesourcery.com/gnu_toolchains/arm/}, 1997-.
CodeSourcery's bug database is not accessible to the public; the mailing
list for their ARM compiler is archived at
\url{http://www.codesourcery.com/archives/arm-gnu}.

14, 29: Kaner, Cem, Falk, Jack, and Nguyen, Hung Q., \emph{Testing
Computer Software,} Second Edition, ISBN: 0471358460, Wiley 1999.

15: Beizer, Boris, \emph{Software System Testing and Quality Assurance,}
ISBN: 0442213069, Van Nostrand 1984.

16, 21, 36: GCC C-Torture Test Suite:
\url{http://gcc.gnu.org/install/test.html}.

17, 50: Cygwin \url{http://www.cygwin.com}, 1995-.

18: Perl (Practical Extraction and Report Language), \emph{Programming
Perl,} Third Edition, Wall, Larry et al., ISBN: 0-596-000278, O'Reilly
\& Associates 2000, \url{http://www.perl.org}, 1987-.

19: Blitz++ C++ Class Library for Scientific Computing,
\url{http://oonumerics.org/blitz}, 1996-.

22: Boost C++ Libraries, \url{http://www.boost.org}, 1998-.

23: Delta, a tool for test failure minimization, Wilkerson, Daniel and
McPeak, Scott, \url{http://delta.tigris.org}, 2003-5. Based on {[24,
Zeller]}. See also {[59, Open Source Quality Project]}.

24: Zeller, A.: ``Yesterday, my program worked. Today, it does not.
Why?'', \emph{Software Engineering - ESEC/FSE'99: 7th European Software
Engineering Conference}, ISSN 0302-9743, volume 1687 of Lecture Notes in
Computer Science, pp. 253-267, 1999.

25: Harbison, Samuel P., and Steele, Guy L., \emph{C: A Reference
Manual,} Fifth Edition, ISBN: 013089592X,
\url{http://www.careferencemanual.com/}, Prentice Hall 2002.

26: ARMulator: ARM Emulator, GDB Version:
\url{http://sources.redhat.com/cgi-bin/cvsweb.cgi/src/sim/arm/?cvsroot=src},
1999-. See also the results of the documentation `info gdb Target',
subsection `Target Commands', paragraph `target sim', in versions of GDB
with ARM emulator support. This requires building with the
`-\/-target=arm-none-elf' option; see {[}58, GDB Bug 1884{]}.

27: Macintosh OSX, Apple Computer, Inc.,
\url{http://developer.apple.com/referencelibrary/MacOSX}, 2001-.

28: ARM Application Binary Interface (ARM ABI),
\url{http://www.arm.com/products/DevTools/ABI.html}, 2001-.

33: CodeSourcery 2005Q3-2, Release announcement for the 2005Q3-2 version
of the CodeSourcery GNU Toolchain for ARM Processors,
\url{http://www.codesourcery.com/archives/arm-gnu-announce/msg00006.html},
2005.

37, 55: Lindig, Christian, ``Random Testing of the Translation of C
Function Calls,''
\url{http://www.st.cs.uni-sb.de/~lindig/src/quest}.
\emph{Proceedings of the Sixth International Workshop on
Automated Debugging}, ISBN 1-59593-050-7, Association for
Computing Machinery 2005.

38: Tydeman, Fred, C99 FPCE Test Suite, \url{http://www.tybor.com/readme.1st},
1995-2006.

39: Kahan, William, Sumner, Thos, et al., Paranoia Floating Point Test,
\url{http://www.netlib.org/paranoia/paranoia.c}, 1983-5.

40: Lischner, Ray, \emph{C++ In a Nutshell,} ISBN: 0-596-00298-X,
O'Reilly 2003. Source code at \url{http://examples.oreilly.com/cplsian}.

41: STLport Standard Library Project \url{http://www.stlport.org}, 1997-

42: lcc, A Retargetable Compiler for ANSI C,
\url{http://www.cs.princeton.edu/software/lcc/}; described in \emph{A
Retargetable C Compiler: Design and Implementation,} Hanson, David R.
and Fraser, Christopher W., ISBN: 0-8053-1670-1, Benjamin/Cummings
Publishing 1995.

43, 57: Fraser, Christopher and Hanson, David, \emph{A Retargetable C
compiler: Design and Implementation,} ISBN: 0-8053-1670-1,
Benjamin/Cummings Publishing, 1995.

44: Kernighan, Brian W. and Ritchie, Dennis M., \emph{The C Programming
Language,} Second Edition, ISBN: 0131103628, Prentice Hall 1988.

45: Small Device C Compiler (SDCC), Dutta, Sandeep et al.,
\url{http://sdcc.sourceforge.net/}, 1999-.

46: Freely Distributable Mathematics Library (FDLIBM), Sun Microsystems,
Inc., \url{http://www.netlib.org/fdlibm/readme}, 1993.

48: Dominus, Mark Jason, ``Good Advice and Maxims for Programmers,''
\url{http://jwenet.net/notebook/2005/1036.html}, 2002.

52: DejaGnu \url{http://www.gnu.org/software/dejagnu/}, 1993-.

53: Burgess, C.J. , ``Bibliography for Automatic Test Data Generation
for Compilers'', comp.compilers,
\url{http://compilers.iecc.com/comparch/article/93-12-064}, 1993.

54: Bailey, Mark W. and Davidson, Jack W., ``Automatic Detection and
Diagnosis of Faults in Generated Code for Procedure Calls'', IEEE
Transactions on Software Engineering, volume 29, issue 11, 2003. An
abstract is available online, at
\url{http://csdl.computer.org/comp/trans/ts/2003/11/e1031abs.htm}, as is an
earlier version of the full paper,
\url{http://big-oh.cs.hamilton.edu/~bailey/pubs/techreps/TR-2001-1.pdf}

56: Aho, Alfred V., Sethi, Ravi, and Ullman, Jeffrey D.,
\emph{Compilers: Principles, Techniques and Tools,} ISBN: 0201100886,
Addison Wesley 1986.

58: GNU Project Debugger (GDB) Bugzilla Database:
\url{http://sources.redhat.com/cgi-bin/gnatsweb.pl}.

59: Open Source Quality Project \url{http://osq.cs.berkeley.edu/}.
\end{raggedright}
\end{hangparas}

\bigskip
\begin{small}
Copyright © 2002-2006, ACCESS Systems Americas, Inc. PalmSource, Palm OS
and Palm Powered, and certain other trade names, trademarks and logos
are trademarks which may be registered in the United States, France,
Germany, Japan, the United Kingdom and other countries, and are either
owned by PalmSource, Inc. or its affiliates, or are licensed exclusively
to PalmSource, Inc. by Palm Trademark Holding Company, LLC. All other
brands, trademarks and service marks used herein are or may be
trademarks of, and are used to identify other products or services of,
their respective owners. All rights reserved.
\end{small}

\section{[End of main article]}

\subsection{[Bibliography, in original alphabetical order]}

\begin{hangparas}{.25in}{1}
\begin{raggedright}
Aho, Alfred V., Sethi, Ravi, and Ullman, Jeffrey D., \emph{Compilers:
Principles, Techniques and Tools,} ISBN: 0201100886, Addison Wesley
1986.

Apple Bug Database \url{https://bugreport.apple.com/}, free registration
required.

Apogee Software, Inc., \url{http://www.apogee.com/compilers.html}, 1988-.

ARM Application Binary Interface (ARM ABI),
\url{http://www.arm.com/products/DevTools/ABI.html}, 2001-.

ARM Developer Suite (ADS), version 1.2,
\url{http://www.arm.com/support/ads_faq.html}, 2001-2004.

ARMulator: ARM Emulator, GDB Version:
\url{http://sources.redhat.com/cgi-bin/cvsweb.cgi/src/sim/arm/?cvsroot=src},
1999-. See also the results of the documentation `info gdb Target',
subsection `Target Commands', paragraph `target sim', in versions of GDB
with ARM emulator support. This requires building with the
`-\/-target=arm-none-elf' option; see {[}GDB Bug 1884{]}.

Bailey, Mark W. and Davidson, Jack W., ``Automatic Detection and
Diagnosis of Faults in Generated Code for Procedure Calls'', IEEE
Transactions on Software Engineering, volume 29, issue 11, 2003. An
abstract is available online, at
\url{http://csdl.computer.org/comp/trans/ts/2003/11/e1031abs.htm}, as is an
earlier version of the full paper,
\url{http://big-oh.cs.hamilton.edu/~bailey/pubs/techreps/TR-2001-1.pdf}

Beizer, Boris, \emph{Software System Testing and Quality Assurance,}
ISBN: 0442213069, Van Nostrand 1984.

Blitz++ C++ Class Library for Scientific Computing,
\url{http://oonumerics.org/blitz}, 1996-.

Boost C++ Libraries, \url{http://www.boost.org}, 1998-.

Burgess, C.J. , ``Bibliography for Automatic Test Data Generation for
Compilers'', comp.compilers,
\url{http://compilers.iecc.com/comparch/article/93-12-064}, 1993.

\emph{C++98 ISO/IEC Standard,} ANSI/ISO 14882:1998(E),
\url{http://www.techstreet.com/cgi-bin/detail?product\_id=1143945}, 1998.

\emph{C99 ISO/IEC Standard,} INCITS/ISO/IEC 9899-1999, second edition,
\url{http://webstore.ansi.org/ansidocstore/product.asp?sku=INCITS\%2FISO\%2FIEC+9899-1999},
1999. Commonly referred to as ``ANSI C.''

CodeSourcery GNU Toolchain for ARM Processors,
\url{http://www.codesourcery.com/gnu_toolchains/arm/}, 1997-.
CodeSourcery's bug database is not accessible to the public; the mailing
list for their ARM compiler is archived at
\url{http://www.codesourcery.com/archives/arm-gnu}.

CodeSourcery 2005Q3-2, Release announcement for the 2005Q3-2 version of
the CodeSourcery GNU Toolchain for ARM Processors,
\url{http://www.codesourcery.com/archives/arm-gnu-announce/msg00006.html},
2005.

Cygwin \url{http://www.cygwin.com}, 1995-.

DejaGnu \url{http://www.gnu.org/software/dejagnu/}, 1993-.

Delta, a tool for test failure minimization, Wilkerson, Daniel and
McPeak, Scott, \url{http://delta.tigris.org}, 2003-5. Based on {[Zeller]}.
See also {[Open Source Quality Project]}.

Dominus, Mark Jason, ``Good Advice and Maxims for Programmers,''
\url{http://jwenet.net/notebook/2005/1036.html}, 2002.

Edison Design Group, Inc., \url{http://www.edg.com/cpp.html}, 1991-.

Fraser, Christopher and Hanson, David, \emph{A Retargetable C compiler:
Design and Implementation,} ISBN: 0-8053-1670-1, Benjamin/Cummings
Publishing, 1995.

Freely Distributable Mathematics Library (FDLIBM), Sun Microsystems,
Inc., \url{http://www.netlib.org/fdlibm/readme}, 1993.

GNU Compiler Collection (GCC) \url{http://gcc.gnu.org}, 1987-.

GCC Bugzilla Database: \url{http://gcc.gnu.org/bugzilla}. See also
\url{http://pobox.com/~flash/FlashsOpenSourceBugReports.html}
for an ongoing list of bugs found with the techniques presented here.

GCC C-Torture Test Suite: \url{http://gcc.gnu.org/install/test.html}.

GNU Project Debugger (GDB), \url{http://www.gnu.org/software/gdb}, 1988-.

GDB Bugzilla Database:
\url{http://sources.redhat.com/cgi-bin/gnatsweb.pl}.

Griffith, Arthur, \emph{GCC: The Complete Reference,} ISBN: 0072224053,
McGraw-Hill 2002.  (Errata: 
\url{http://pobox.com/~flash/errata/GCC_The_Complete_Reference_Errata.html})

Harbison, Samuel P., and Steele, Guy L., \emph{C: A Reference Manual,}
Fifth Edition, ISBN: 013089592X, \url{http://www.careferencemanual.com/},
Prentice Hall 2002.

Kahan, William, Sumner, Thos, et al., Paranoia Floating Point Test,
\url{http://www.netlib.org/paranoia/paranoia.c}, 1983-5.

Kaner, Cem, Falk, Jack, and Nguyen, Hung Q., \emph{Testing Computer
Software,} Second Edition, ISBN: 0471358460, Wiley 1999.

Kernighan, Brian W. and Ritchie, Dennis M., \emph{The C Programming
Language,} Second Edition, ISBN: 0131103628, Prentice Hall 1988.

lcc, A Retargetable Compiler for ANSI C,
\url{http://www.cs.princeton.edu/software/lcc/}; described in \emph{A
Retargetable C Compiler: Design and Implementation,} Hanson, David R.
and Fraser, Christopher W., ISBN: 0-8053-1670-1, Benjamin/Cummings
Publishing 1995.

Lindig, Christian, ``Random Testing of the Translation of C Function
Calls'', \url{http://www.st.cs.uni-sb.de/~lindig/src/quest}.
\emph{Proceedings of the Sixth International Workshop on}

\emph{Automated Debugging}, ISBN 1-59593-050-7, Association for
Computing Machinery 2005.

Lischner, Ray, \emph{C++ In a Nutshell,} ISBN: 0-596-00298-X, O'Reilly
2003. Source code at \url{http://examples.oreilly.com/cplsian}.

Open Source Quality Project \url{http://osq.cs.berkeley.edu/}.

Palm OS® Developer Suite:
\url{http://www.palmos.com/dev/tools/dev_suite.html}, free registration
required, 2004-.

Palm OS Cobalt®, \url{http://www.palmos.com/dev/tech/oses/cobalt60.html},
2003-.

Perl (Practical Extraction and Report Language), \emph{Programming
Perl,} Third Edition, Wall, Larry et al., ISBN: 0-596-000278, O'Reilly
\& Associates 2000, \url{http://www.perl.org}, 1987-.

Small Device C Compiler (SDCC), Dutta, Sandeep et al.,
\url{http://sdcc.sourceforge.net/}, 1999-.

STLport Standard Library Project \url{http://www.stlport.org}, 1997-

Tydeman, Fred, C99 FPCE Test Suite, \url{http://www.tybor.com/readme.1st},
1995-2006.

Zeller, A.: ``Yesterday, my program worked. Today, it does not. Why?'',
\emph{Software Engineering - ESEC/FSE'99: 7th European Software
Engineering Conference}, ISSN 0302-9743, volume 1687 of Lecture Notes in
Computer Science, pp. 253-267, 1999.

\end{raggedright}
\end{hangparas}

\end{document}